\documentclass[twocolumn,prl,twocolumn,superscriptaddress]{revtex4}

\usepackage[latin9]{inputenc}
\usepackage{color}
\usepackage{amsmath}
\usepackage{amssymb}
\usepackage{graphicx}
\usepackage[unicode=true,
 bookmarks=true,bookmarksnumbered=false,bookmarksopen=false,
 breaklinks=false,pdfborder={0 0 1},backref=false,colorlinks=true]
 {hyperref}
\hypersetup{
 linkcolor=magenta, urlcolor=blue, citecolor=blue, pdfstartview={FitH}, hyperfootnotes=false, unicode=true}

\makeatletter
\@ifundefined{textcolor}{}
{%
 \definecolor{BLACK}{gray}{0}
 \definecolor{WHITE}{gray}{1}
 \definecolor{RED}{rgb}{1,0,0}
 \definecolor{GREEN}{rgb}{0,1,0}
 \definecolor{BLUE}{rgb}{0,0,1}
 \definecolor{CYAN}{cmyk}{1,0,0,0}
 \definecolor{MAGENTA}{cmyk}{0,1,0,0}
 \definecolor{YELLOW}{cmyk}{0,0,1,0}
}

\usepackage{amsfonts}\usepackage{tabularx}\usepackage{dcolumn}\usepackage{bm}\usepackage{graphicx}\usepackage{epstopdf}

\hypersetup{urlcolor=blue}

\makeatother

\begin{document}

\title{Machine learning topological states}

\author{Dong-Ling Deng}
\affiliation{Condensed Matter Theory Center and Joint Quantum Institute, Department
of Physics, University of Maryland, College Park, MD 20742-4111, USA}
\author{Xiaopeng Li} 
\affiliation{State Key Laboratory of Surface Physics, Institute of Nanoelectronics and Quantum Computing, and Department of Physics, Fudan University, Shanghai 200433, China}
\affiliation{Collaborative Innovation Center of Advanced Microstructures, Nanjing 210093, China} 
\affiliation{Condensed Matter Theory Center and Joint Quantum Institute, Department
of Physics, University of Maryland, College Park, MD 20742-4111, USA} 

\author{S. Das Sarma}
\affiliation{Condensed Matter Theory Center and Joint Quantum Institute, Department
of Physics, University of Maryland, College Park, MD 20742-4111, USA}

\begin{abstract} 
Artificial neural networks and machine learning  have now reached a new era after several decades of improvement where applications are to explode in many fields of science, industry, and technology. Here, we use artificial neural networks to study an intriguing phenomenon in quantum physics--- the topological phases of matter. 
We find that certain topological states, either symmetry-protected or with intrinsic topological order, can be represented with classical artificial neural networks.   
This is demonstrated by using three concrete spin systems, the one-dimensional (1D) symmetry-protected topological cluster state and the 2D and 3D toric code states with  intrinsic topological orders. For all three cases we show rigorously that the topological ground states can be represented by short-range neural networks in an \textit{exact} and \textit{efficient} fashion---the required number of hidden neurons is as small as the number of physical spins and the number of parameters scales only \textit{linearly} with the system size.   For the 2D toric-code model, we find that the proposed short-range neural networks can describe the excited states with abelain anyons and their nontrivial mutual statistics as well. In addition, by using reinforcement learning we show that  neural networks are capable of finding the topological ground states of non-integrable Hamiltonians with strong interactions and studying their topological phase transitions. Our results demonstrate explicitly the exceptional power of neural networks in describing topological quantum states, and at the same time provide valuable guidance to machine learning of topological phases in generic lattice models.

\end{abstract}
\maketitle

\section{INTRODUCTION}
Machine learning, grown out of the quest for artificial intelligence,
is one of today's most active fields across disciplines with vast applications
ranging from fundamental research in cheminformatics, biology, and cosmology to quantitative social sciences~\cite{Jordan2015Machine,Lecun2015Deep}. Within physics, machine learning
techniques have recently  been  introduced for gravitational wave analysis~\cite{Rahul2013Application,Abbott2016Observation},
black hole detection~\cite{Pasquato2016Detecting},
and material design~\cite{Kalinin2015Big}. 
More recently, these techniques have been utilized to improve numerics in studying phase transitions in conventional systems~\cite{Arsenault2015Machine,Carrasquilla2017Machine,Wang2016Discovering,
Torlai2016Learning,Broecker2017Machine,Ch2016Machine}. 


In general, for a quantum system to fully describe a many-body state
requires a huge amount of information due to the exponential scaling of 
the Hilbert space dimension \cite{Gharibian2014Quantum}. Yet,
typical physical states may only access a tiny corner
of the Hilbert space and in principle could be represented in a 
much restricted subspace of reduced dimension. 
For example, the area law entangled quantum states~\cite{Hastings2007Area} 
can be efficiently represented in terms
of matrix product states (MPS) \cite{Fannes1992Finitely,Perez2007Matrix,Schollwock2011Density} or tensor-network states in general \cite{Gu2009Tensor,Orus2014Practical,Verstraete2008Matrix}, 
which grant efficient numeric algorithms to solve 
complex quantum many-body problems,  e.g., 
DMRG (density-matrix renormalization group) method~\cite{White1992Density,Schollwock2005TheDMRG,Schollwock2011Density,Vidal2003Efficient}.  
Recently, an artificial-neural-network approach has been proposed as a completely different route to implement ``Hilbert-space reduction"~\cite{Carleo2016Solving}, which opens up a  new thrust of machine-learning based  algorithms to simulate quantum many-body systems. 

Although the existence of a neural network representation of arbitrary quantum
states is assured by mathematical theorems \cite{Le2008Representational,Hornik1991Approximation,Kolmogorov1963Representation}, 
how the required classical resources scale  is {\it unknown}. 
 In particular, it is unknown whether the topological states, of crucial relevance to condensed matter physics~\cite{Wen2004Quantum,chen2012symmetry,Chiu2016Classification,qi2011topological,hasan2010colloquium}, 
can be expressed by neural networks efficiently. These are fundamental questions standing in the way of applying machine learning techniques to topological quantum phases of matter. 

In this paper, we find that topological states can be represented with artificial neural networks in an {\it exact} and {\it efficient} way from one to three dimensions. Our study is based on exact construction of neural network representations for  the cluster state in 1D, and toric codes in 2D and 3D. Strikingly, for the toric-code states we find that they can be represented precisely by short-range neural networks, despite their long-range entanglement properties associated with topological orders.  
Moreover, we find that  our constructed neural networks can be used to describe anyon braidings and their nontrivial mutual statistics as well. 
In order to show that neural networks can also describe topological states in more generic non-integrable systems, we numerically study a 1D interacting Hamiltonian through reinforcement learning. We demonstrate that  neural networks are capable of finding the corresponding topological ground states  and studying the phase transition (from a smmetry protected topological phase to a ferromagnetic phase) when tuning the interaction strength. Our results provide valuable guidance and data resources to the applications of machine-learning techniques to  topological phases in condensed matter physics.

\section{Artificial-neural-network representation} 

\begin{figure}[htp]
\includegraphics[width=0.45\textwidth]{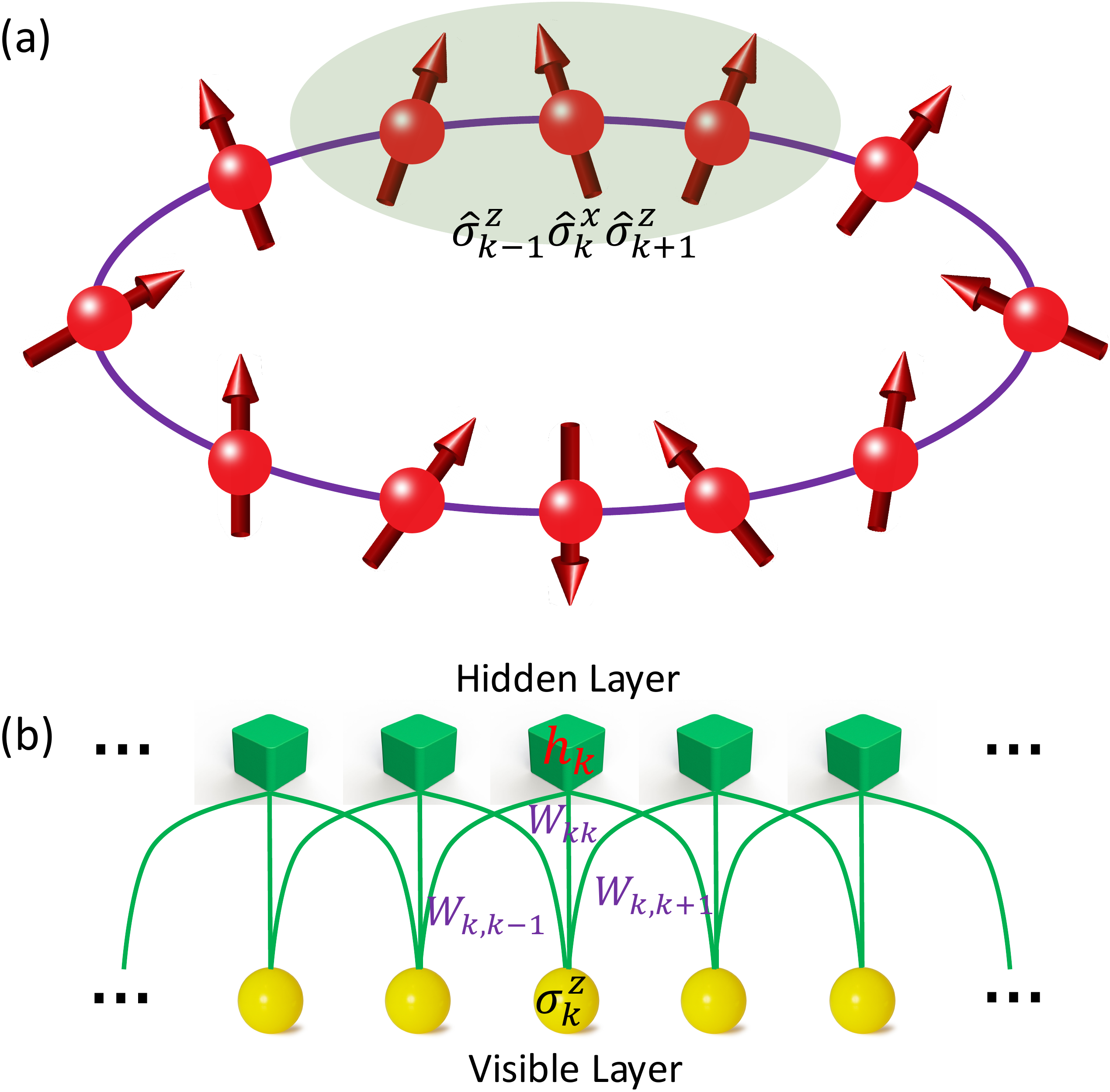}

\caption{The 1D symmetry-protected topological cluster state. (a) A pictorial illustration of the 1D cluster Hamiltonian
(see Eq. (\ref{eq:1DHam})) with the periodic boundary condition.
The shaded region represents a prototypic three-body interaction (a
stabilizer) at an arbitrary site $k$. (b) A short-range neural network
representation of the 1D topological cluster state. The yellow balls
(blue cubes) denote the visible (hidden) artificial neurons. \label{fig:1D-graph-state.}}
\end{figure}

To begin
with, let us first outline the artificial-neural-network representation
of quantum states, which has recently been introduced by Carleo and Troyer 
in solving many-body problems via machine learning ideas \cite{Carleo2016Solving}.
Considering a quantum system with $N$ spins $\Xi=(\sigma_{1},\sigma_{2},\cdots\sigma_{N})$,
we use the restricted Boltzmann machine (RBM), which is a stochastic
artificial neural network with widespread applications \cite{Hinton2006Reducing,Salakhutdinov2007Restricted,Larochelle2008Classification,Amin2016Quantum,Mehta2014Exact,Adachi2015Application},
to describe the many-body wave-function $\Phi(\Xi)$. We focus on spin-$1/2$ quantum systems.
The RBM contains two layers \cite{Le2008Representational}, one visible layer of $N$ nodes corresponding
to the physical spins, and the other a hidden layer of $M$ auxiliary
classical spin variables $h_{1},\cdots,h_{M}$ (see Fig. \ref{fig:1D-graph-state.}(b)
for an 1D example). An artificial-neural-network quantum state (ANNQS)
has the form \cite{Carleo2016Solving}: 
$
\Phi_{M}(\Xi;\Omega) =  \sum_{\{h_{k}\}}e^{\sum_{k}a_{k}\sigma_{k}^{z}+\sum_{k'}b_{k'}h_{k'}+\sum_{kk'}W_{kk'}h_{k}\sigma_{k'}^{z}},
$
where $\{h_{k}\}=\{-1,1\}^{M}$ denotes the possible configurations
of $M$ hidden auxiliary spins and the weights $\Omega=(a_{k},b_{k'},W_{kk'})$
are parameters needed to train to best represent the many-body quantum
state. ANNQS should be taken as a variational state and for a given
$\Phi_{M}(\Xi;\Omega)$, the actual quantum many-body state $|\Phi\rangle$
is understood as (up to an irrelevant normalization constant) $|\Phi\rangle=\sum_{\Xi}\Phi_{M}(\Xi;\Omega)|\Xi\rangle$,
similar to the Laughlin-like representation of the exact resonating-valence-bond
ground state of the Haldane-Shastry model \cite{Haldane1988Exact,Shastry1988Exact}.

While the  representability theorems  
guarantee the existence of ANNQS to \textit{approximate} arbitrary many-body
state 
\cite{Le2008Representational,Hornik1991Approximation,Kolmogorov1963Representation}, 
such existence would not be practically useful if an exponential (in system size) number of neurons are required. 
 Moreover, given a specific quantum system, there is so far \textit{no} systematic way to write down its wave function in terms of ANNQS. 
It is thus desirable to construct {\it exact}  ANNQS for non-trivial quantum many-body systems. 
In this work, we first introduce a {\it further restricted} RBM (FRRBM), where the hidden neurons connect only locally to the visible neurons. This further restriction builds in locality property for the RBM, which enables us to
 use ANNQS
to represent certain topological states exactly and efficiently (in
the sense that both the number of neurons and the number of weight
variables are linear in system size). We give three explicit examples, one for
the SPT cluster state in 1D and the other two for the toric code states with
intrinsic topological orders in 2D and 3D, respectively.  For more generic cases, we release this further restriction and use  RBM to numerically solve the topological ground state of a non-integrable  
Hamiltonian and study the corresponding topological phase transition via reinforcement learning techniques.

\section{1D SPT cluster state}

We consider the following
Hamiltonian defined on a 1D lattice with periodic boundary condition
(Fig. \ref{fig:1D-graph-state.}a):
\begin{eqnarray}
H_{\text{cluster}} & =- & \sum_{k=1}^{N}\hat{\sigma}_{k-1}^{z}\hat{\sigma}_{k}^{x}\hat{\sigma}_{k+1}^{z},\label{eq:1DHam}
\end{eqnarray}
where $\hat{\sigma}^{z}$ and $\hat{\sigma}^{x}$ are Pauli matrices
and $N$ denotes the system size. Throughout this paper, we use $\hat{\bullet}$ for  operators (e.g., $\hat{\sigma}^{z}$ and $\hat{\sigma}^{x}$
) and $\sigma^{z}$ for classical variables ($\sigma^{z}=\pm1$).
The Hamiltonian $H_{\text{cluster}}$ has two $Z_{2}$ symmetries
corresponding to $\hat{\sigma}^{z,y}\rightarrow-\hat{\sigma}^{z,y}$ for either even- or odd-indexed sites, 
and its ground state is a topological state protected by $Z_{2}\times Z_{2}$
symmetry \cite{Son2012Topological}, analogous to the Haldane phase
of the spin-1 chain \cite{Haldane1983Nonlinear,Affleck1987Rigorous,affleck1988valence}. 
Due to the stabilizer nature of the Hamiltonian, the ground state $|G\rangle$ obeys
\begin{eqnarray}
\hat{\sigma}_{k-1}^{z}\hat{\sigma}_{k}^{x}\hat{\sigma}_{k+1}^{z}|G\rangle & = & |G\rangle,\quad\forall k.\label{eq:1DEqG}
\end{eqnarray}
In the context of quantum information and computation, 
this state $|G\rangle$ 
is called a cluster state \cite{Briegel2001Persistent} or more generally
a graph state \cite{Hein2004Multparty}. It has important applications
in measurement-based quantum computation \cite{Hein2004Multparty,Raussendorf2003Measurement,Nielsen2006Cluster}.
We note that a variant of $H_{\text{cluster}}$
has also been studied recently in the context of many-body localization,
where the symmetry protected topological phase is shown to persist even 
to highly excited eigenstates due to localization protection
\cite{Bahri2015Localization}. 

Here we construct an exact artificial neural network representation for the cluster state. 
Before we provide details of the construction and derivation, we first outline the main ideas. Generally speaking, for an arbitrary quantum many-body state, there is no known method to recast it in the ANNQS form. This is a challenging task intuitively because there are ``infinitely many'' possibilities to choose the network structure and weight variables. For the cases studied in this paper, our main idea is, rather than to use the general network with long-range connections, to choose a special case with only short-range connections.  Using the fact that the RBM does not contain intra-layer connections, we can factor out the hidden variables and rewrite $\Phi_{M}(\Xi;\Omega)$ in a product form. We then explore the constraints on the ground state (such as Eq. (\ref{eq:1DEqG})) to build up a series of nonlinear equations and simplify these equations by canceling all the equal factors on both sides. We solve these equations by recasting them to an optimization problem to numerically find out the nonzero parameters. If no solution can be found, we choose another network with more nonzero undetermined parameters and repeat the procedure until a solution is found. We validate the exactness of the solution in a rigorous way by analytically checking that all the equations are satisfied.

In the ANNQS representation, we have $|G\rangle=\sum_{\Xi}\Phi_{M}(\Xi;\Omega)|\Xi\rangle$. 
Then Eq.~\ref{eq:1DEqG} gives, 
\begin{eqnarray}
\hat{\sigma}_{k-1}^{z}\hat{\sigma}_{k}^{x}\hat{\sigma}_{k+1}^{z}\sum_{\Xi}\Phi_{M}(\Xi;\Omega)|\Xi\rangle & = & \sum_{\Xi}\Phi_{M}(\Xi;\Omega)|\Xi\rangle.\quad\label{eq:1DGsNNS}
\end{eqnarray}
Our aim is to design a network and solve the weight variables $\Omega$
to satisfy the above equation for all $k$. Noting that the operator
$\hat{\sigma}_{k}^{x}$ will flip spin on site $k$ and $\hat{\sigma}_{j}^{z}|\Xi\rangle=\sigma_{j}^{z}|\Xi\rangle$,
we can reduce Eq. (\ref{eq:1DGsNNS}) to 
\begin{eqnarray}
\Phi_{M}(\Xi;\Omega)\sigma_{k-1}^{z}\sigma_{k+1}^{z} & = & \Phi_{M}(\Xi,\sigma_{k}^{z}\rightarrow-\sigma_{k}^{z};\Omega).\label{eq:spinflipeq}
\end{eqnarray}
We can 
 explicitly factor out the hidden variables and rewrite $\Phi_{M}(\Xi;\Omega)$
in a product form \cite{Carleo2016Solving} $\Phi_{M}(\Xi;\Omega)=\prod_{k=1}^{N}e^{a_{k}\sigma_{k}^{z}}\prod_{k'=1}^{M}\Gamma_{k'}(\Xi)$,
with $\Gamma_{k'}(\Xi)=2\cosh(b_{k'}+\sum_{l}W_{k'l}\sigma_{l}^{z})$.
As a result, Eq. (\ref{eq:spinflipeq}) can be reduced to
\begin{eqnarray}
\sigma_{k-1}^{z}\sigma_{k+1}^{z}e^{2a_{k}\sigma_{k}^{z}}\prod_{l=1}^{M}\Gamma_{k'}(\Xi)=\prod_{l=1}^{M}\Gamma_{k'}(\Xi,\sigma_{k}^{z}\rightarrow-\sigma_{k}^{z}), \label{eq:EssEq} 
\end{eqnarray}
which is the essential equation to determine the actual neural network. An exact representation
of $|G\rangle$ requires Eq. (\ref{eq:EssEq}) to be satisfied for
all spin configurations $\Xi$ and all $k$. This gives us a series
of (on the order of $\sim N2^{N}$) highly nonlinear equations, which
is hard to solve directly for general neural networks. With a finite
number of neurons, it is {\it unknown} whether a solution even exists. We will thus not attempt to solve the problem for a general
neural network. Instead, we will work in an alternate direction.
By noting that 
the cluster state $|G\rangle$ has translational invariance and is short-range 
entangled (it has area law entanglement entropy~\cite{Hastings2007Area}), 
it is reasonable to assume that the RBM representation for $|G\rangle$ has a  further restricted form with the inter-layer neurons only locally coupled. This assumption will be validated by showing analytically  that the final proposed ANNQS satisfy Eq. (\ref{eq:EssEq}). A simple  FRRBM is shown in Fig. \ref{fig:1D-graph-state.}(b), with the weight
parameters chosen to be
\begin{eqnarray}
a_{k}  =  0,\; b_{k}=ib,\;
W_{kj}  =  \begin{cases}
i\omega_{k-j}, & \text{if }|k-j|=1\\
0 & \text{otherwise}
\end{cases}.\label{eq:useTransSymm}
\end{eqnarray}

Here we choose the parameters to be purely imaginary to convert the hyperbolic cosine functions to cosine functions. 
By plugging Eq. (\ref{eq:useTransSymm})
into Eq. (\ref{eq:EssEq}) and canceling all the equal factors at both sides, we
arrive at 
\begin{eqnarray}
\sigma_{k-1}^{z}\sigma_{k+1}^{z}\Lambda_{k-1}\Lambda_k\Lambda_{k+1}=\Lambda_{k-1}^-\Lambda_k^-\Lambda_{k+1}^-,\label{eq:Coseq}
\end{eqnarray}
where $\Lambda_p=\cos(b+\omega_{1}\sigma_{p-1}^{z}+\omega_{0}\sigma_{p}^{z}+\omega_{-1}\sigma_{p+1}^{z})$ ($p=k-1,k,k+1$) and $\Lambda_p^-=\Lambda_p(\sigma^z_k\rightarrow -\sigma^z_k)$.
Eq.(\ref{eq:Coseq}) should be satisfied for any spin configurations
of $\Xi_{\text{sub}}=(\sigma_{k-2}^{z},\sigma_{k-1}^{z},\sigma_{k}^{z},\sigma_{k+1}^{z},\sigma_{k+2}^{z})$,
giving rise to a set of $2^5$ equations.
Directly solving these highly-nonlinear equations is still daunting.
But we can recast it as an optimization problem. We define a function
$f(b,\omega_{1},\omega_{0},\omega_{-1})=\sum_{\Xi_{\text{sub}}}(L_{\Xi_{\text{sub}}}-R_{\Xi_{\text{sub}}})^{2}$
where $L_{\Xi_{\text{sub}}}$ ($R_{\Xi_{\text{sub}}}$) denotes the
left (right) side of Eq. (\ref{eq:Coseq}) with configuration $\Xi_{\text{sub}}$,
and then numerically minimize $f$ over $(b,\omega_{1},\omega_{0},\omega_{-1})$.
A zero minimum of $f$ corresponds to a desired solution to Eq. (\ref{eq:Coseq}), and we find 
$(b,\omega_{1},\omega_{0},\omega_{-1})=\frac{\pi}{4}(1,2,3,1)$. 
The corresponding ANNQS representation of $|G\rangle$ is now obtained, 
\begin{eqnarray}
\Phi_{M}(\Xi;\Omega) & = & \sum_{\{h_{k}\}}e^{\frac{i\pi}{4}\sum_{k}h_{k}(1+2\sigma_{k-1}^{z}+3\sigma_{k}^{z}+\sigma_{k+1}^{z})}.\label{eq:main1dres}
\end{eqnarray}
This gives a compact neural network representation of the cluster state whose  
 number of nonzero parameters scales linearly with the system size($\sim4N$). We stress that, although we have resorted  to numerics to solve the highly nonlinear equations,  our final FRRBM-based representation 
is \textit{exact}, in the sense that the corresponding ANNQS satisfies  Eq. (\ref{eq:1DEqG}) exactly.  In fact, the exactness can be verified analytically after some straightforward calculations. A key observation is that $\Lambda_p=-\Lambda_p(\sigma^z_{p-1}\rightarrow -\sigma^z_{p-1})$ for $(b,\omega_{1},\omega_{0},\omega_{-1})=\frac{\pi}{4}(1,2,3,1)$. Thus, we have $\Lambda_{k+1}=-\Lambda_{k+1}^-$ and the Eq. (\ref{eq:Coseq}) reduces to $\sigma_{k-1}^{z}\sigma_{k+1}^{z}\Lambda_{k-1}\Lambda_k=-\Lambda_{k-1}^-\Lambda_k^-$. This can be obtained from the fact that $\sigma_{k+1}\Lambda_k=\Lambda_k^-$ and $\sigma_{k-1}\Lambda_{k-1}=-\Lambda_{k-1}^-$. 

\begin{figure}
\includegraphics[width=0.471\textwidth]{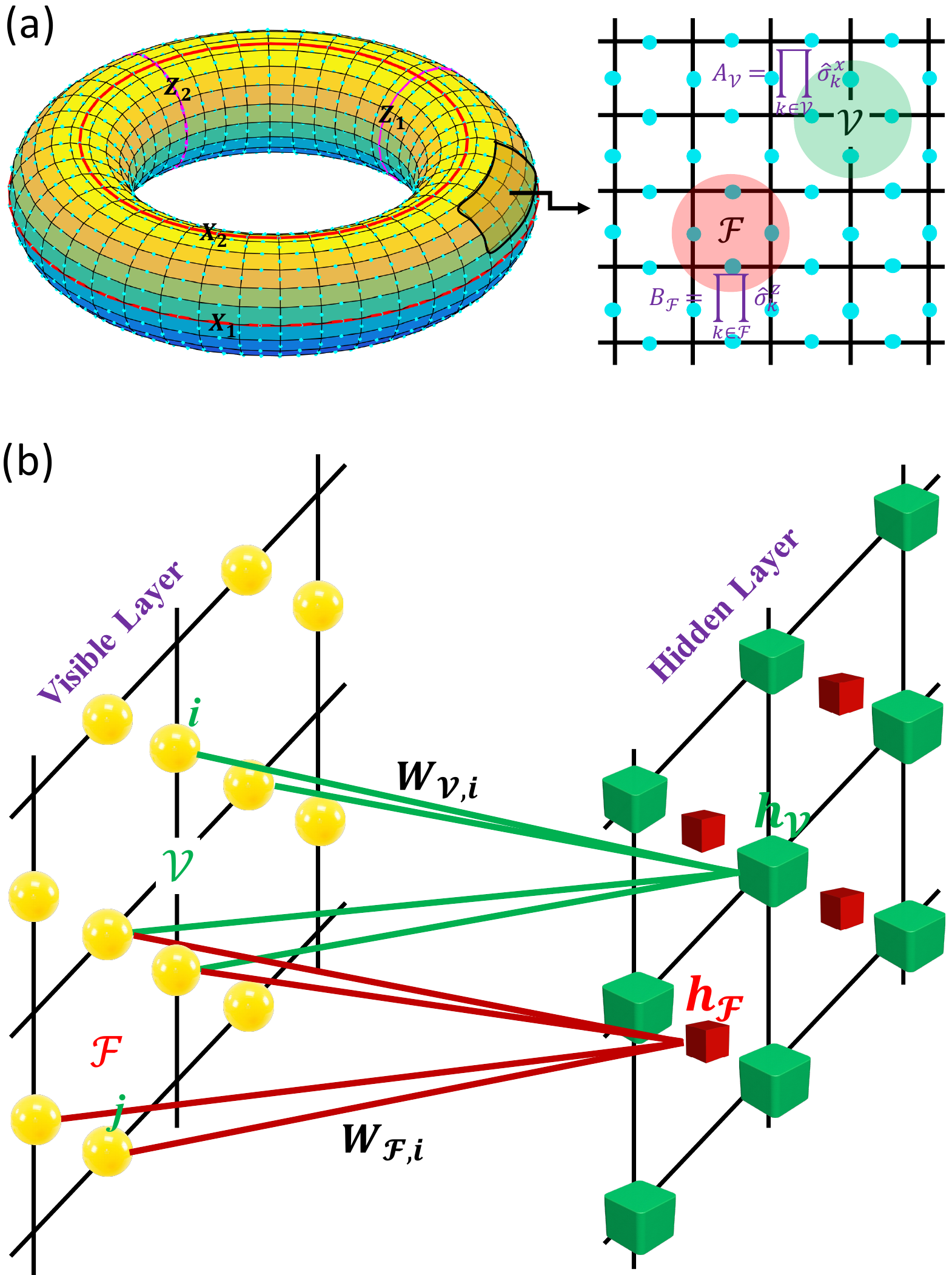}

\caption{The 2D toric code Hamiltonian and a neural-network representation.
(a) An illustration of $H_{\text{2D}}$ (see Eq. (\ref{eq:Kitaevmodel})).
The shaded green (red) region depicts a prototypic vertex (face) operator
at vertex $\mathcal{V}$ (face $\mathcal{F}$). The Hamiltonian is
defined as a summation of these four-body operators with
an extra minus sign in front. The four non-contractible loops (labeled
by $X_{1}$, $X_{2}$, $Z_{1}$ and $Z_{2}$, respectively) correspond
to four nontrivial string operators that define the commutation relations
of the two logic qubits living in the topologically protected ground
subspace \cite{kitaev2003fault,hu2009proposed}. (b) A neural-network
representation of one of the ground state of $H_{\text{2D}}$ with intrinsic topological order.
The yellow balls stand for the visible artificial neurons corresponding
to the physical spins. The green (red) cubes denotes the hidden neurons
corresponding to the vertices (faces). \label{fig:2dtorriccode}}
\end{figure}

\section{2D Kitaev toric code states} 
As our second example, we study the 2D toric code model \cite{kitaev2003fault}, which was introduced by Kitaev \cite{kitaev2003fault}
in the context of topological quantum computation \cite{nayak2008non}
and quantum error correction \cite{Lidar2013Quantum}. This model gives  the simplest
and most well studied spin liquid ground state  that has an intrinsic $Z_2$ topological order~\cite{Read1991LargeN,Wen1991Meanfield}.
Considering a $L\times L$ square lattice with the periodic boundary
condition (a 2D torus $\mathbb{T}^{2}$)  with each edge of
the lattice attached a qubit (Fig. \ref{fig:2dtorriccode}). We have 
$N=2L^{2}$ qubits in total. For each vertex $\mathcal{V}$ (face
$\mathcal{F}$) (see Fig.~\ref{fig:2dtorriccode}(a)) we define a vertex (face) operator $A_{\mathcal{V}}=\prod_{k\in\mathcal{V}}\hat{\sigma}_{k}^{x}$
($B_{\mathcal{F}}=\prod_{k\in\mathcal{F}}\hat{\sigma}_{k}^{z}$),
which are also called stabilizers in quantum error correction language.
The model Hamiltonian reads:
\begin{eqnarray}
H_{\text{2D}} & = & -\sum_{\mathcal{V}\in\mathbb{T}^{2}}A_{\mathcal{V}}-\sum_{\mathcal{F}\in\mathbb{T}^{2}}B_{\mathcal{F}}.\label{eq:Kitaevmodel}
\end{eqnarray}
This model is exactly solvable since all the four-body operators
in $H_{\text{2D}}$ commute with each other. It can be interpreted
as a particular Ising lattice gauge theory \cite{Kogut1979Anintroduction}
with an abelian $Z_{2}$ gauge group \cite{Kitaev2010Topological}.
Its ground state is four-fold degenerate, a signature of  intrinsic
topological order. The low-energy excitations are abelian anyons with
nontrivial mutual statistics \cite{kitaev2003fault}. Because of its
fundamental importance in the studies of quantum computing and topological
phases of matter, the  toric code  has attracted tremendous
interest in both theory \cite{Duan2003Controlling,hu2009proposed,Kitaev2010Topological,Aguado2008Creation,Hamma2005Ground}
and experiment \cite{Du2007Experimental,Gladchenko2009Superconducting,Lu2009Demonstrating,
Pachos2009Revealing,Zhong2016Emulating}. 
The ground state of the model satisfies
$
B_{\mathcal{F}}|G_{\rm toric}^{(\text{2D})} \rangle=|G_{\rm toric}^{(\text{2D})} \rangle
$ and $A_{\mathcal{V}}|G_{\rm toric}^{(\text{2D})}\rangle=|G _{\rm toric}^{(\text{2D})} \rangle$. We note that $|G_{\rm toric}^{(\text{2D})}\rangle $ has an efficient tensor-network representation \cite{Csahinouglu2014Characterizing,Verstraete2006Criticality}.

Here, we show that  $|G_{\rm toric}^{(\text{2D})}\rangle $ has an 
\textit{exact} and \textit{efficient}  neural-network representation. We present the result
 and the verification here and provide the more involved construction details in 
the appendixes. 
As illustrated in Fig.~\ref{fig:2dtorriccode}, the artificial neural network representation of the toric code ground state, 
$|G_{\rm toric}^{(\text{2D})}\rangle=\sum_{\Xi}\Phi_{M}(\Xi)|\Xi\rangle$, is given by
\begin{eqnarray}
 \Phi_{M}(\Xi;\Omega) 
  =  \sum_{\{h_{\mathcal{V}},h_{\mathcal{F}}\}}e^{\frac{i\pi}{2}\sum\limits_{\mathcal{V}}h_{\mathcal{V}}\sum\limits_{j\in\mathcal{V}}\sigma_{j}^{z}+\frac{i\pi}{4}\sum\limits_{\mathcal{F}}h_{\mathcal{F}}\sum\limits_{k\in\mathcal{F}}\sigma_{k}^{z}}. \label{Toric2DNNWRep}
\end{eqnarray}

To verify the the exact nature of our solution, as done in the 1D example, we can factor out  
the hidden neurons,  $\Phi_{M}(\Xi;\Omega)=\prod_{\mathcal{V}}\cos(\frac{\pi}{2}(\sum_{j\in\mathcal{V}}\sigma_{j}^{z}))\prod_{\mathcal{F}}\cos(\frac{\pi}{4}(\sum_{j\in\mathcal{F}}\sigma_{j}^{z}))$.
Noting that $\cos(\frac{\pi}{4}\sum_{j\in\mathcal{F}}\sigma_{j}^{z})\prod_{k\in\mathcal{F}}\sigma_{k}^{z}=\cos(\frac{\pi}{4}\sum_{j\in\mathcal{F}}\sigma_{j}^{z})$,
the equation $B_{\mathcal{F}}|G_{\rm toric}^{(\text{2D})} \rangle=|G_{\rm toric} ^{(\text{2D})}\rangle$ can be easily
verified. In order to verify $A_{\mathcal{V}}|G_{\rm toric} ^{(\text{2D})}\rangle=|G_{\rm toric} ^{(\text{2D})}\rangle$,
we need to show 
$\Phi_{M}(\Xi )=\Phi_{M}(\Xi,\sigma_{j}^{z}\rightarrow-\sigma_{j}^{z},\forall j\in\mathcal{V})$.
This actually follows from two observations about the consequence of flipping spins belonging to a vertex 
$\mathcal{V}$, which are  the  sign-change of all four cosine
factors for the neighboring vertices ${\cal V}$s and the sign-preservation of the
product of the four cosine factors for the neighboring faces ${\cal F}$s.\\

\begin{figure}
\includegraphics[width=0.46\textwidth]{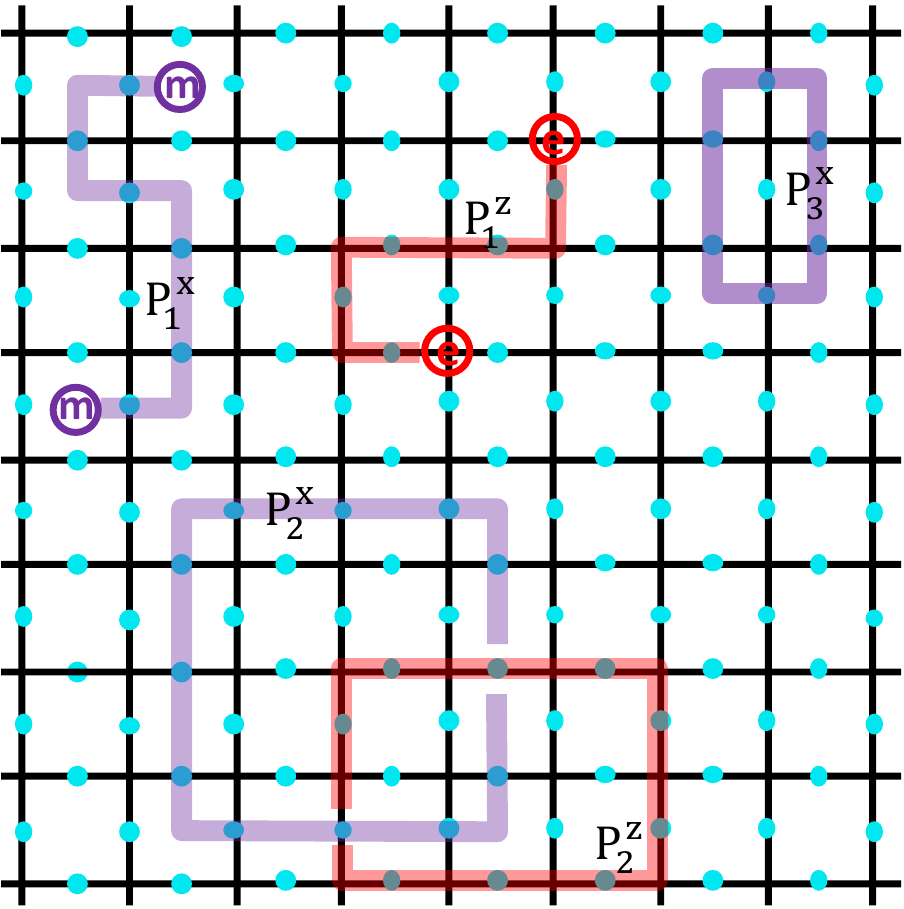}

\caption{String operators and nontrivial mutual statistics.
A pair of x-type quasiparticles (also called ``magnetic vortices") living on the faces can be created by acting the string operator $S^x_{\text{P}^\text{x}_1}=\prod_{j\in \text{P}^\text{x}_1}\hat{\sigma}^x_j$ on the ground state $|G_{\rm toric}^{(\text{2D})}\rangle$. Similary, we can also create a pair of z-type quasiparticles (``electric charges") that live on the vertices by the string operator $S^z_{\text{P}^{\text{z}}_1}=\prod_{j\in \text{P}^{\text{z}}_1}\hat{\sigma}^z_j$. For a closed contractable path (such as $\text{P}^x_3$), two quasiparticles of the same type meet with each other and will fuse to the vacuum.  If two closed loops of different types form a nontrivial link (such as the Hopf link formed by $\text{P}^\text{x}_2$ and $\text{P}^\text{z}_2$), the many-body states will aquire a global phase factor $-1$ due to the nontrivial mutual statistics between the x-type and z-type quasiparticles. As shown in the main text and the Appendix B, we find that all the quantum states evovled in these process can be described precisely and efficiently by further-restricted restricted Boltzmann machines.  \label{fig:2danyons}}
\end{figure}

More interestingly, we notice that the proposed FRRBM can precisely describe the excited states with abelian anyons and their nontrivial mutual statistics as well. In the 2D Kitaev toric code model, there are two types of anyons \cite{kitaev2003fault}: z-type quasiparticles (or ``electric charges") living on the vertices and x-type quasiparticles (or ``magnetic vortices") living on the faces, which can be created in pairs by the string operators $S^z_{\text{P}^\text{z}}=\prod_{j\in \text{P}^{\text{z}}}\hat{\sigma}^z_j$  and $S^x_{\text{P}^\text{x}}=\prod_{j\in \text{P}^{\text{x}}}\hat{\sigma}^x_j$, respectively. The x-type (z-type) quasiparticles exist at the endpoints of the path  $\text{P}^\text{x}$ ($\text{P}^\text{z}$) and they can be moved around by extending or shortening $\text{P}^\text{x}$ ($\text{P}^\text{z}$) (see Fig. \ref{fig:2danyons} for an illustration). When two quasiparticles of the same type meet, they will annihilate each other to vacuum (a fusion process). In addition, if we braid a x-type particle around a z-type particle and then annihilate them with their partners, we will obtain a global phase $-1$ due to the nontrivial mutual statistics between the x- and z-type particles. This process correspond to applying the string operators of two closed paths that are linked together (such as the Hopf link formed by $\text{P}^x_2$ and $\text{P}^z_2$ in Fig. \ref{fig:2danyons}) to the ground states (see Appendix B). 

We now show how to describe all the processes discussed above in the FRRBM framework (see Appendix B for details). We present two key observations: (i) applying a string operator $S^x_{\text{P}^\text{x}}$  is equivalent to flipping all signs of the weight parameters associated with the visible neurons living on the path $\text{P}^\text{x}$; (ii) applying  $S^z_{\text{P}^\text{z}}$  is equivalent to adding hidden neurons along $\text{P}^\text{z}$, with each of them connecting only to the corresponding visible neuron. Based on these observations, a FRRBM description of the excited states with abelian anyons follows directly. For instance, let us consider an excited state $|\Psi_{\text{P}^\text{x}_1}\rangle$  with a pair of x-type particles at the ends of $\text{P}^\text{x}_1$ (Fig.\ref{fig:2danyons}): $|\Psi_{\text{P}^\text{x}_1}\rangle=S^x_{\text{P}^\text{x}_1}|G_{\rm toric}^{(\text{2D})}\rangle$. To represent $|\Psi_{\text{P}^\text{x}_1}\rangle$ by a FRRBM, we simply flip all signs of the parameters (as specified in Eq. (\ref{Toric2DNNWRep})) that are associated with the visible neurons living on $\text{P}^\text{x}_1$. More explicitly, $|\Psi_{\text{P}^\text{x}_1}\rangle$ has the following exact RBM representation
%
\begin{eqnarray}
 \Phi_{M} 
  &=&  \sum_{\{h_{\mathcal{V}},h_{\mathcal{F}}\}}\exp[\frac{i\pi}{2}\sum\limits_{\mathcal{V}}h_{\mathcal{V}}(\sum\limits_{j\in\mathcal{V};j\notin \text{P}^\text{x}_1}\sigma_{j}^{z}-\sum\limits_{j\in\mathcal{V};j\in \text{P}^\text{x}_1}\sigma_{j}^{z}) \nonumber \\
  &+& \frac{i\pi}{4}\sum\limits_{\mathcal{F}}h_{\mathcal{F}}(\sum\limits_{k\in\mathcal{F};k\notin \text{P}^\text{x}_1}\sigma_{k}^{z}-\sum\limits_{k\in\mathcal{F};k\in \text{P}^\text{x}_1}\sigma_{k}^{z})].  \label{Toric2DAnnyons}
\end{eqnarray}
Similarly, we can write down the exact RBM expressions for other excited states with different number of x- or z-particles. In the Appendix B, we verify that these neural-network states are indeed eigenstates of $H_{\text{2D}}$ with the corresponding quasiparticles at the desired locations.   We also show that the braiding process and  the resulting nontrivial mutual statistics can be described precisely using the RBM language as well. This result explicitly shows that neural networks are capable of exactly (and not just efficiently) describing exotic states with abelian anyons. 
In the future, we hope that it can inspire more studies on the applications of machine-learning ideas in numerically simulating anyons (both abelian and non-abelian) and their braidings in strongly correlated systems.

\begin{figure}
\includegraphics[width=0.471\textwidth]{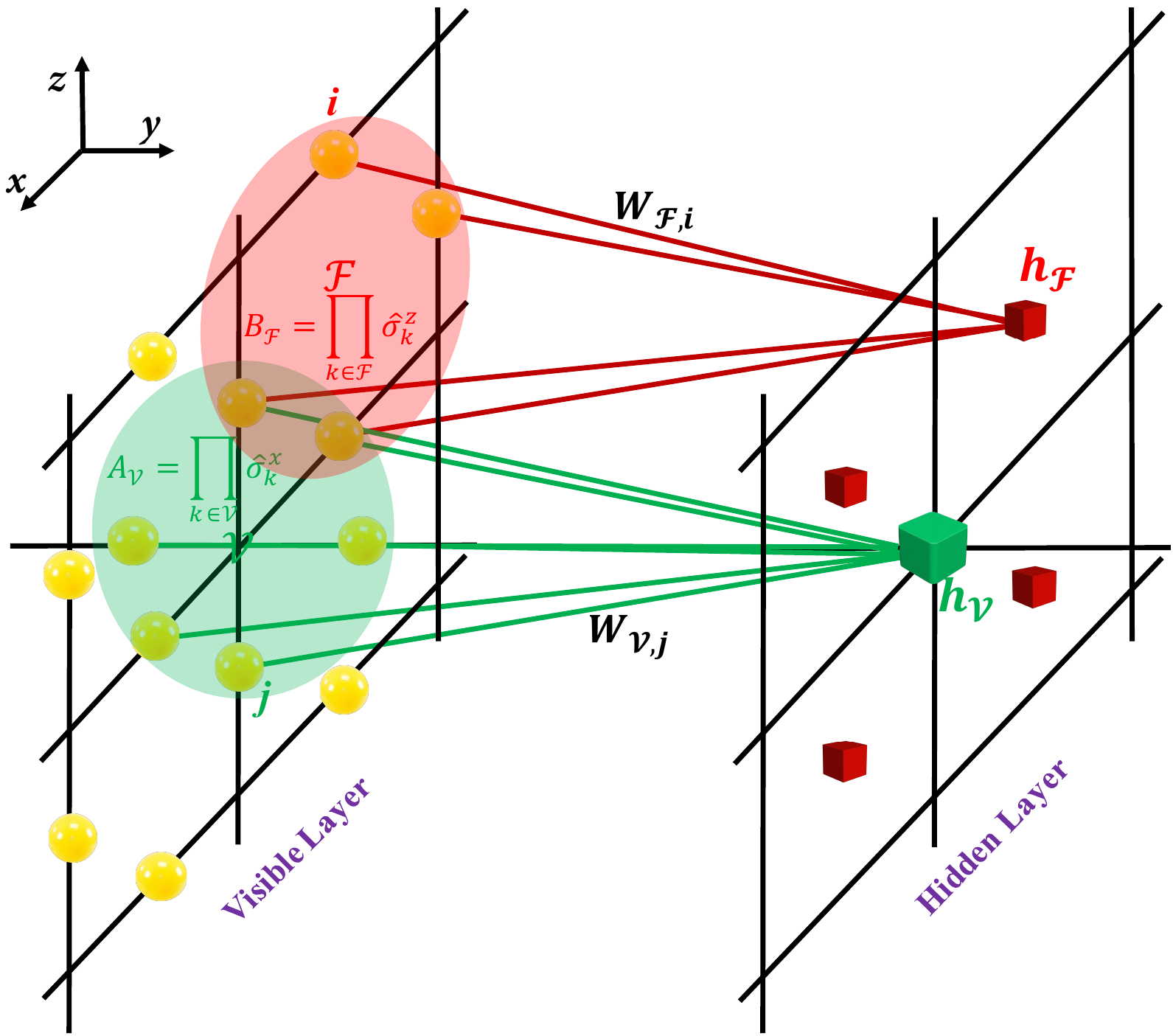}

\caption{  A neural-network representation for the 3D toric code states.
Similar to the 2D model, the 3D Hamiltonian is also defined as a summation of vertex and face operators (depicted by the shaded green and red regions, respectively) with an extra minus sign in front. But, for the 3D case each vertex operator will be a product of six, rather than four, $\hat{\sigma}^x$ operators. The ground states of $H_{\text{3D}}$ can also be represented by restricted Boltzmann machines, where the hidden neurons (denoted by green or red cubes) only connect to their nearest visible neurons.  The weight parameters $W_{\mathcal{F},i}$ and $W_{\mathcal{V},j}$ can be chosen the same as in the 2D case. 
\label{fig:3dtorriccode}}
\end{figure}

\section {3D toric code states} 

The above 2D toric code model has a natural 3D generalization defined on a simple  cubic lattice with the periodic boundary condition (a 3D torus $\mathbb{T}^3$) \cite{Hamma2005String,Castelnovo2008Topological}: 
\begin{eqnarray}
H_{\text{3D}} & = & -\sum_{\mathcal{V}\in\mathbb{T}^{3}}A_{\mathcal{V}}-\sum_{\mathcal{F}\in\mathbb{T}^{2}}B_{\mathcal{F}},\label{eq:3Dmodel}
\end{eqnarray}
where $A_{\mathcal{V}}=\prod_{k\in\mathcal{V}}\hat{\sigma}_{k}^{x}$
and $B_{\mathcal{F}}=\prod_{k\in\mathcal{F}}\hat{\sigma}_{k}^{z}$ are the corresponding vertex and face operators, respectively. This model is  exactly solvable and has been a paradigmatic model  of topological order in 3D. It features both closed-string and closed-membrane condensation \cite{Hamma2005String}. 
At finite temperatures, it can exhibit "classical" topological order (in the sense that the topological entropy comes only from the plaquette degrees of freedom and the system looks like purely classical) up to a transition temperature $T_c$ (see Ref. \cite{Castelnovo2008Topological} for details).  This is in sharp contrast to the 2D case where the total topological entropy vanishes in the thermodynamic limit at any nonzero temperature with the topological order being argued to be fragile  \cite{Castelnovo2008Topological}. 

Using similar methods as discussed in the 2D case  above, we can also find an exact and efficient FRRBM representation for the ground states $|G_{\text{toric}}^{\text{(3D)}}\rangle$ of $H_{\text{3D}}$. In fact, we find that one can use the same weight parameters as specified by Eq. (\ref{Toric2DNNWRep}) to represent $|G_{\text{toric}}^{\text{(3D)}}\rangle$ (see Fig. \ref{fig:3dtorriccode}). The exactness of this representation can be verified by showing that the equations $B_{\mathcal{F}}|G_{\rm toric}^{(\text{3D})} \rangle=|G_{\rm toric} ^{(\text{3D})}\rangle$ and $A_{\mathcal{V}}|G_{\rm toric}^{(\text{3D})} \rangle=|G_{\rm toric} ^{(\text{3D})}\rangle$ are satisfied. The first equation is straightforward and the second equation follows from the fact that flipping spins belonging to a vertex will not change the sign of the product of the nearest eighteen cosine factors affected by $A_{\mathcal{V}}$. We mention that the FRRBM can also describe the low-energy excited states of $H_{\text{3D}}$  (obtained by applying different string or membrane operators on the ground state)  and their mutual statistics, analogous to the 2D case.


It is worthwhile to point out that, although the above results have rigorously established that both 2D and 3D toric code states have an exact and efficient FRRBM representation, it still requires substantial future efforts to find out the necessary and sufficient conditions for a generic quantum many-body state with intrinsic topological order to manifest a FRRBM representation. Future studies of this problem would facilitate the applications of machine learning techniques in investigating topological phases of matter. Conversely, such studies may also provide valuable insights for understanding why certain machine learning algorithms are so powerful, similar to the heuristic example of how we understand the power of DMRG algorithm from the perspective of MPS representation. We hope that our work demonstrating exact representations of 1D cluster SPT states as well as 2D and 3D toric codes using FRRBM would inspire future research into the generic applications of machine learning techniques to topological phenomena (including the simulation of abelian and nonabelian anyons) and to the understanding of the specific physical features underlying machine learning algorithms making them suitable for understanding entanglement features in topological models.

\begin{figure}
\includegraphics[width=0.486\textwidth]{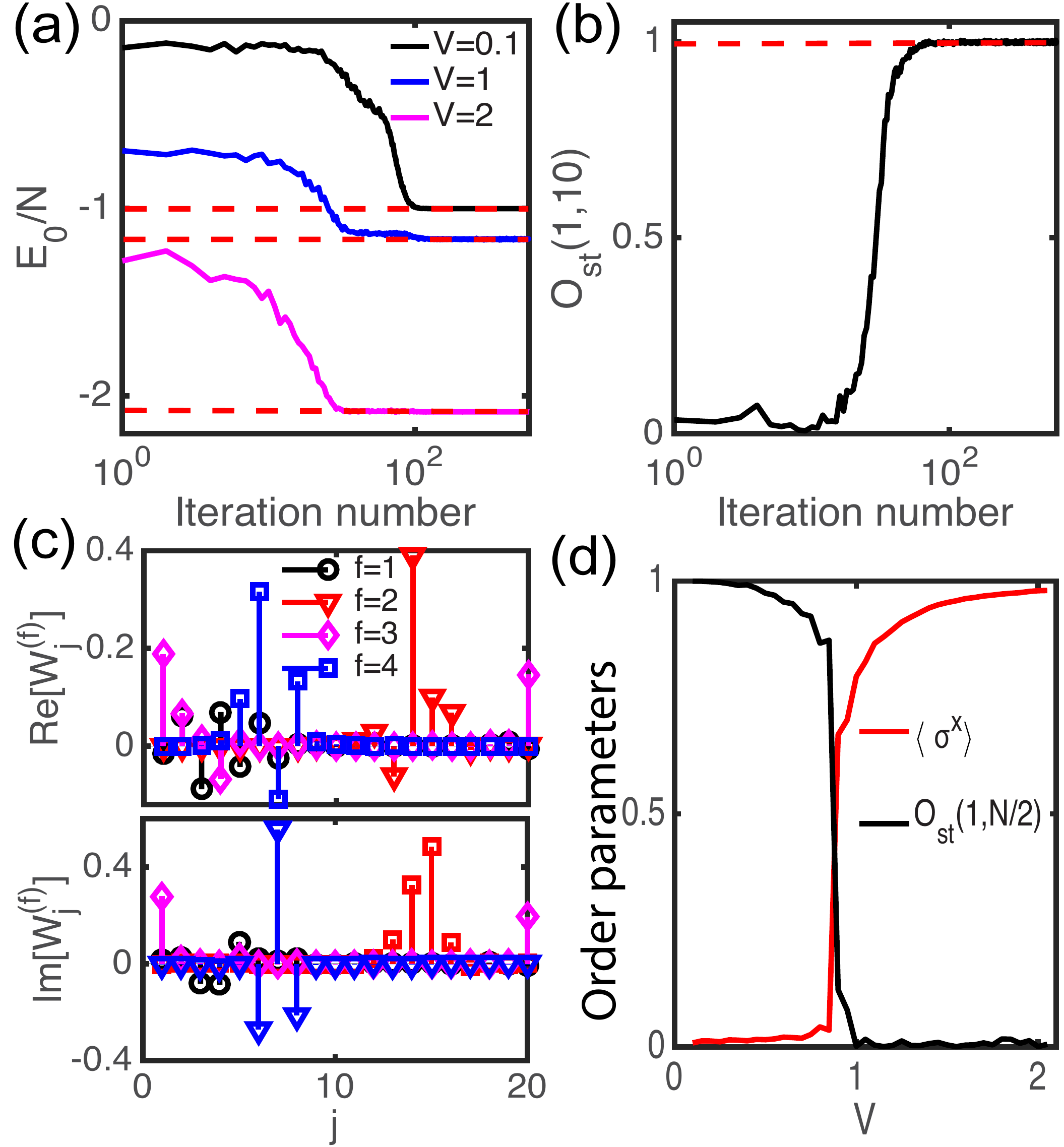}

\caption{Reinforcement learning of SPT states and topological phase transitions. (a) The variational ground-state energy density as a function of the iteration number of the learning process. As the iteration number increases, the energy converges smoothly to the exact values (denoted by the dashed red lines). (b) The convergence of the string-order parameter as the increasing of the iteration number for $V=0.1$. the red dashed line denotes the exact value.  (c) The real and imaginary parts of the learned feature maps for representing the ground state of $H_{\text{1D}}$ with a restricted Boltzmann machine for $V=0.1$. In (a), (b), and (c), the lattice size is $N=20$, $h_x=0.02$ and the hidden-unit density $\gamma=4$. (d) Reinforceent learning of a topological phase transition. As we increasing $V$, the system will go through a phase transition (at $V\sim0.9$) from a smmetry protected topological phase to a ferromagnetic phase.  Here the lattice size is $N=50$ and the other omitted parameters are chosen the same as in (b).
\label{fig:ReInLSPT1D}}
\end{figure}

\section {Reinforcement learning of SPT phases and phase transitions}

In above discussions, we have given analytical results on representing topological states with neural networks for some exactly solvable models. In this section, we consider more generic cases and show, through numerical reinforcement learning \cite{Carleo2016Solving,Sorella2007Weak}, that RBM is capable of finding the topological ground states of non-integrable interacting Hamiltonians and studying their topological phase transitions. To this end, we add an interaction term and a magnetic field into the 1D SPT cluster Hamiltonian to break its integrability. We consider the following Hamiltonian:
\begin{eqnarray}
H_{\text{1D}}=-H_{\text{cluster}}-h_x\sum_{k=1}^N\sigma^x_k-V\sum_{k=1}^N\sigma^x_k\sigma^x_{k+1},
\end{eqnarray}
Where $h_x$ and $V$ denote the strengths of the magnetic field and interaction, respectively. It is obvious that $H_{\text{1D}}$ maintains the same $Z_2\times Z_2$ symmetry. In the limit $h_x, V \rightarrow 0$, the ground state of $H_{\text{1D}}$ is a SPT state, same to that of  $H_{\text{cluster}}$ (up to a local gauge transformation). We use a string-order parameter $O_{\text{st}}(j,k)=\langle \sigma^z_j\sigma^y_{j+1}(\prod_{i=j+2}^{k-2}\sigma^x_i)\sigma^y_{k-1}\sigma^z_k\rangle$ to character the topological nature of this SPT state \cite{Bahri2014Detecting}. Whilst in the other limit $V \rightarrow \infty$ (we fix $h_x=0.02$ for simplicity), the ground state will be a ferromagnetic state, breaking the $Z_2\times Z_2$ symmetry. There should be a phase transition point at some critical value $V=V_c$  where the system goes  from the SPT phase to the ferromagnetic phase. Here, by using reinforcement learning we show that  RBM is capable of efficiently and faithfully representing the SPT ground state of $H_{\text{1D}}$ and pinning down the critical value $V_c$.

Noting that $H_{\text{1D}}$ has a lattice translational symmetry, we can use it to reduce the number of variational parameters, and for integer hidden-variable density ($\gamma\equiv M/N=1,2,\cdots$), the weight matrix takes the form of feature filters $W^{(f)}_j$ with $f\in [1,\gamma]$ an integer number \cite{Carleo2016Solving}. In Fig.\ref{fig:ReInLSPT1D}(a) and (b), we plot the ground state energy density and the string-order parameter obtained via reinforcement learning and compare the RBM result with that from exact diagonalization (ED), for small system sizes. We see that as the iteration number of the learning process increases, both of them converge smoothly to their corresponding exact values. As in Ref. \cite{Carleo2016Solving}, we quantify the accuracy of the trained RBM by the relative error on the ground-state energy $\epsilon\equiv|E_0^{RBM}-E_0^{ED}|/E_0^{ED}$. In Fig. \ref{fig:ReInLSPT1D}(c), we plot the feature maps after a typical reinforcement learning process with $\gamma=4$ and $N=20$, where the accuracy $\epsilon\sim 10^{-3}$.  One can systematically further inprove this accuracy by increaing $\gamma$ and the number of iterations. In Fig. \ref{fig:ReInLSPT1D}(d), we show the order parameters obtained through RBM for a larger system size (which is far beyond the capability of the ED technique).  We find that both the magnetization (characterizing the ferromagnetic phase) and the string-order parameter have sharp jumps around $V_c\sim 0.9$, indicating a  phase transition from the SPT phase  to the ferromagnetic phase across this value. 
We mention that the reinforcement learning techniques may also be used to study topological states and phase transitions in higher dimensions. We leave this for future studies.

\section {CONCLUSION and Discussion}  

In summary, we have demonstrated, both analytically and numerically, that quantum topological states (both symmetry protected and  intrinsic) can be efficiently represented by classical artificial neural networks.
We have constructed exact representations for SPT states (the 1D cluster states)  and  intrinsic topologically ordered states (2D and 3D toric code states), by using  the FRRBM method.   
For all cases, the number of neurons in the hidden layer of the RBM is equal to the number of physical spins, and the number of nonzero weight variables scales only linearly with the system
size. For the toric-code models, we show that the proposed FRRBM is also capable of  describing the excited states with abelian anyons and theiry nontrivial mutual statistics.
We expect that our construction carries over to other graph states and   the 3D time-reversal SPT phase of bosons with intrinsic  surface topological order \cite{Burnell2014Exactly}. Our method may also generalize to other frustration-free Hamiltonians with translational symmetry, where the ground state is a simultaneous ground state of all local terms \cite{Sattath2016Local}. 
In addition, through numerical reinforcement learning we have also demonstrated that RBM is capable of finding the topological ground states of generic non-integrable Hamiltonians and identifying their topological phase transitions.

Our results manifest the remarkable power of neural networks
in describing and computing exotic quantum states and thus would have far-reaching
implications in the applications of machine learning techniques in
condensed matter physics. 
In practice, our exact results should provide valuable guidance and data resources. 
For instance, our exact results could be used as ``training data'' in supervised learning or the exact parameter values can be used as the initial parameter values for RBM-based reinforcement learning in solving quantum many-body problems. 
In turn, our work may help the study of machine learning itself, especially in the efforts toward understanding why machine learning techniques are surprisingly powerful \cite{Mehta2014Exact,Lin2016Why} from a physical perspective. 
As a first step to connect quantum topology and artificial neuron networks, the present study focused on the simplest network, namely the single-layer model. A straightforward generalization to deep neuron networks is expected to further improve the corresponding representation power, details of which are left for future investigation.



\section*{ACKNOWLEDGMENTS}

We thank Giuseppe Carleo, Roger Melko, Mikhail D. Lukin, Frank Verstraete, Lei Wang, Cheng Fang, Matthias Troyer, Yang-Le Wu, Mohammad Hafezi, Yi Zhang, Eun-Ah Kim, Alexey Gorshkov, and Subir Sachdev for helpful discussions. This work is supported by JQI- NSF-PFC and LPS-MPO-CMTC. XL acknowledge the support by the Start-up Fund of Fudan University.

\renewcommand{\theequation}{A\arabic{equation}}
\setcounter{equation}{0}  
\renewcommand{\thefigure}{A\arabic{figure}}
\setcounter{figure}{0}  

\section*{APPENDIX A: CONSTRUCTING FRRBM FOR THE 2D TORIC CODE STATES}

In this section, we give the details about how we construct the neural network representation of the 2D  toric code states with intrinsic topological order. 

Noting that all these four-body operators in $H_{\text{2D}}$
commute with each other, thus the eigenstates of the Hamiltonian are also the eigenstates of these operators. The ground state satisfies
the following equations:
\begin{eqnarray}
	B_{\mathcal{F}}|G_{\text{toric}}^{\text{2D}}\rangle  &=&  \prod_{j\in\mathcal{F}}\hat{\sigma}_{j}^{z}|G_{\text{toric}}^{\text{2D}}\rangle=|G_{\text{toric}}^{\text{2D}}\rangle,\quad\forall\mathcal{F},\label{eq:BfEq}\\
	A_{\mathcal{V}}|G_{\text{toric}}^{\text{2D}}\rangle & = & \prod_{j\in\mathcal{V}}\hat{\sigma}_{j}^{x}|G_{\text{toric}}^{\text{2D}}\rangle=|G_{\text{toric}}^{\text{2D}}\rangle,\quad\forall\mathcal{V}.\label{eq:Aveq}
\end{eqnarray}
We propose the following neural network state to represent $|G_{\text{toric}}\rangle$: 
\begin{widetext}
\begin{eqnarray}
	\Phi_{M}(\Xi;\Omega)  =  \sum_{\{h_{\mathcal{V}},h_{\mathcal{F}}\}}\exp\Big \{\sum_{k}a_{k}\sigma_{k}^{z}+\sum_{\mathcal{V}}b_{\mathcal{V}}h_{\mathcal{V}}+
	\sum _{\mathcal{F}}b_{\mathcal{F}}h_{\mathcal{F}}+
	\sum_{\mathcal{V}k}W_{\mathcal{V}k}h_{\mathcal{V}}\sigma_{k}^{z}+\sum_{\mathcal{F}k}W_{\mathcal{F}k}h_{\mathcal{F}}\sigma_{k}^{z}\Big \}, \label{eq:NNWSTorric}
\end{eqnarray}
\end{widetext}
where $h_{\mathcal{V}}=\{-1,1\}$ ($h_{\mathcal{F}}=\{-1,1\}$) are
the set of hidden neurons corresponding to the vertices (faces); the
weights $\Omega=(a_{k},b_{\mathcal{V}},b_{\mathcal{F}},W_{\mathcal{V}k},W_{\mathcal{F}k})$
are parameters we need to train. The visible neurons corresponding to the physical spins live on edges of the square lattice.
Throughout this Supplemental Materials, 
the lower-case letters ($k$ or $j$) are used to label individual physical spins (or visible neurons in the restricted Boltzmann machine (RBM) language). For convenience, we also introduce a combined index like (${\cal F}, \mu$), with ${\cal F}$ labeling the faces, and $\mu$ the spins within each face.  
As there is no intra-layer connection in the network, we can rewrite the RBM 
$\Phi_{M}(\Xi;\Omega)$ in a product form:
\begin{eqnarray}
	\Phi_{M}(\Xi;\Omega) & = & \prod_{k=1}^{N}e^{a_{k}\sigma_{k}^{z}}\prod_{\mathcal{V}}\Gamma_{\mathcal{V}}(\Xi)\prod_{\mathcal{F}}\Gamma_{\mathcal{F}}(\Xi),\label{eq:nnwprodf}
\end{eqnarray}
with 
\begin{eqnarray*}
	\Gamma_{\mathcal{V}}(\Xi) & = & 2\cosh[b_{\mathcal{V}}+\sum_{k}W_{\mathcal{V}k}\sigma_{k}^{z}],\\
	\Gamma_{\mathcal{F}}(\Xi) & = & 2\cosh[b_{\mathcal{F}}+\sum_{k}W_{\mathcal{F}k}\sigma_{k}^{z}].
\end{eqnarray*}
To simplify the problem, we introduce a further restriction that the hidden vertex (face) neurons
only connect to the visible neurons belonging to the corresponding
vertex (face) (with a corresponding rule shown in  Fig.2(b) in the main text): 
\begin{eqnarray}
	W_{\mathcal{V}k} & = & 0,\quad\text{if}\;k\notin\mathcal{V},\label{eq:vetex0}\\
	W_{\mathcal{F}k} & = & 0,\quad\text{if}\;k\notin\mathcal{F}.\label{eq:face0}
\end{eqnarray}
We need to find out the weight parameters $\Omega$ so to make Eq.
(\ref{eq:NNWSTorric}) represent the ground state of $H_{\text{2D}}$.
Since the face operators do not involve spin flip, it is easier to
solve Eq. (\ref{eq:BfEq}), and then we get $b_{\mathcal{F}}$ and $W_{\mathcal{F}k}$.
To this end, we plug Eq. (\ref{eq:nnwprodf}) into Eq. (\ref{eq:BfEq})
and obtain
\begin{eqnarray}
	&& \prod _{j\in\mathcal{F}}\sigma_{j}^{z}\prod_{k=1}^{N}e^{a_{k}\sigma_{k}^{z}}\prod_{\mathcal{V}}\Gamma_{\mathcal{V}}(\Xi)\prod_{\mathcal{F}'}\Gamma_{\mathcal{F}'}(\Xi) \\
	&& =\prod_{k=1}^{N}e^{a_{k}\sigma_{k}^{z}}\prod_{\mathcal{V}}\Gamma_{\mathcal{V}}(\Xi)\prod_{\mathcal{F}'}\Gamma_{\mathcal{F}'}(\Xi),\quad\forall\mathcal{F}.\nonumber\label{eq:Beq}
\end{eqnarray}
Canceling all (except $\Gamma_{\mathcal{F}}(\Xi))$ equal factors
on both sides of Eq. (\ref{eq:Beq}), we have 
\begin{eqnarray}
	&& \prod_{j\in\mathcal{F}}\sigma_{j}^{z}  \cosh[b_{\mathcal{F}}+\sum_{\mu=1}^{4}W_{\mathcal{F};(\mathcal{F},\mu)}\sigma_{(\mathcal{F},\mu)}^{z}] \\
	&&= \cosh[b_{\mathcal{F}}+\sum_{\mu=1}^{4}W_{\mathcal{F};(\mathcal{F},\mu)}\sigma_{(\mathcal{F},\mu)}^{z}],\quad\forall\mathcal{F},\nonumber\label{eq:Bafterconcel}
\end{eqnarray}
where we have used $(\mathcal{F},\mu)$ ($\mu=1,2,3,4)$ to denote
the four visible neurons belong to $\mathcal{F}.$ Noting that $\sum_{\mu=1}^{4}\sigma_{(\mathcal{F},\mu)}^{z}=0,\pm2,\pm4$,
it is straightforward to find a solution to Eq. (\ref{eq:Bafterconcel}):
\begin{eqnarray*}
	b_{\mathcal{F}} & = & 0,\;W_{\mathcal{F};(\mathcal{F},\mu)}=\frac{i\pi}{4},\quad\forall\mu,\mathcal{F}.
\end{eqnarray*}

We now turn to the more involved case of solving Eq. (\ref{eq:Aveq})
to obtain $a_{k}$, $b_{\mathcal{V}}$, and $W_{\mathcal{V}k'}$.
Plugging (\ref{eq:nnwprodf}) into Eq. (\ref{eq:Aveq}) and fix $a_{k}=0,\;\forall k$,
we obtain 
\begin{eqnarray*}
	&& \sum_{\varXi}\prod_{\mathcal{V}'}\Gamma_{\mathcal{V}'}(\Xi)\prod_{\mathcal{F}}\Gamma_{\mathcal{F}}(\Xi)  |\varXi;\sigma_{j}^{z}\rightarrow-\sigma_{j}^{z},\forall j\in\mathcal{V}\rangle \\
	 &&=\sum_{\varXi}\prod_{\mathcal{V}'}\Gamma_{\mathcal{V}'}(\Xi)\prod_{\mathcal{F}}\Gamma_{\mathcal{F}}(\Xi)|\varXi\rangle,\;\forall\mathcal{V}.
\end{eqnarray*}
Thus we have 
\begin{eqnarray}
	&&\prod_{\mathcal{V}'}\Gamma_{\mathcal{V}'}(\Xi)\prod_{\mathcal{F}}\Gamma_{\mathcal{F}}(\Xi) 
	=  \prod_{\mathcal{V}'}\Gamma_{\mathcal{V}'}(\Xi;\sigma_{j}^{z}\rightarrow-\sigma_{j}^{z},\forall j\in\mathcal{V}) \nonumber\\	
	&&\quad\times  \prod_{\mathcal{F}}\Gamma_{\mathcal{F}}(\Xi;\sigma_{j}^{z}\rightarrow-\sigma_{j}^{z},\forall j\in\mathcal{V}),\;\forall\mathcal{V}.\label{eq:Avphi}
\end{eqnarray}

\begin{figure}
	\includegraphics[width=0.456\textwidth]{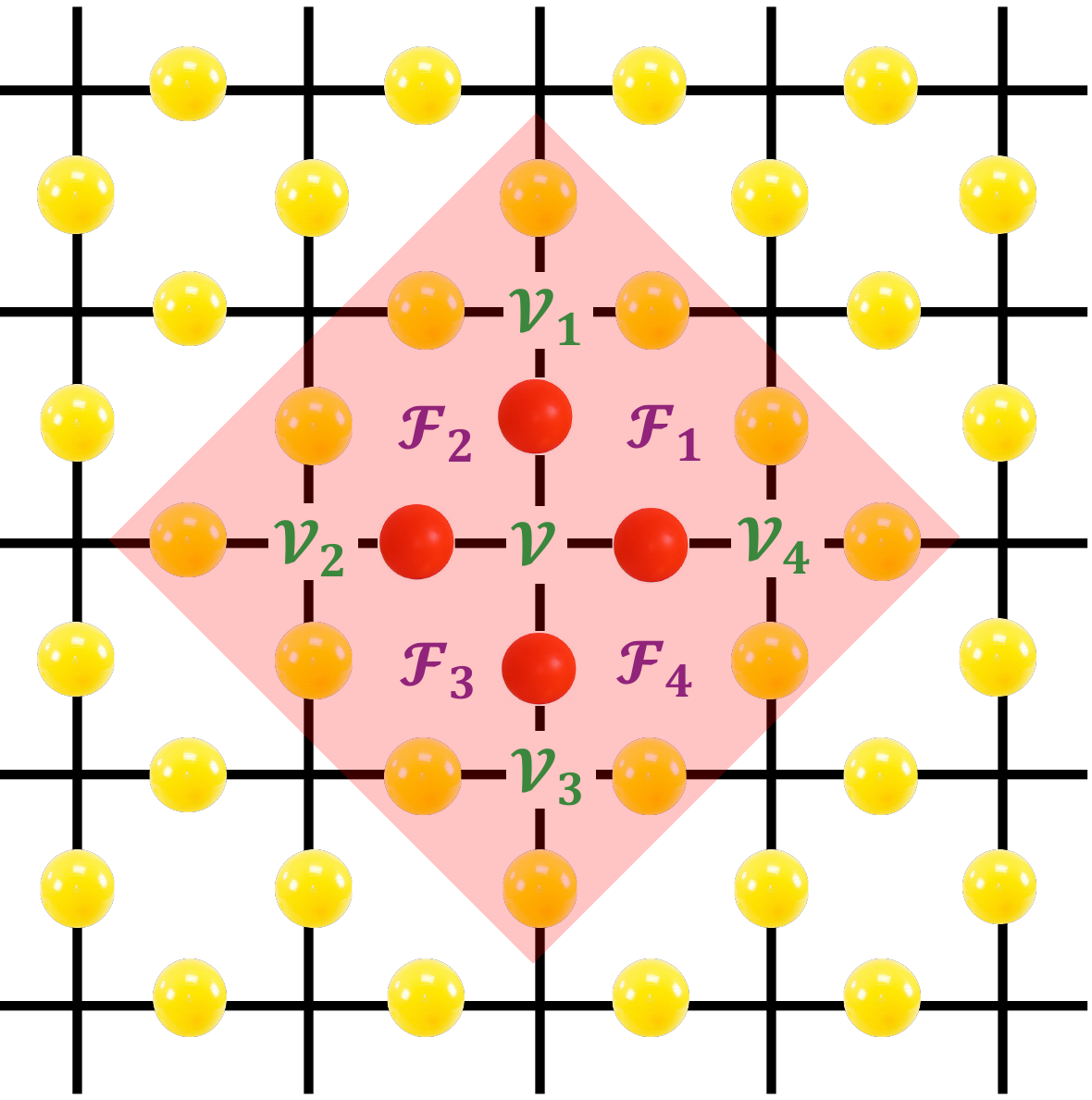}
	
	\caption{Affected region for acting the vertex operator $A_{\mathcal{V}}$.
		$A_{\mathcal{V}}$ flips four spins belonging to $\mathcal{V}$ (denoted
		by the red balls). The shaded region stands for the region that are
		affected by the spin flip. \label{fig:Affected-reurons.}}
\end{figure}

Let us consider spin flips caused by a given vertex $\mathcal{V}$. This corresponding vertex operator $A_{\mathcal{V}}$ only flips four
spins that belong to $\mathcal{V}$. As shown in Fig. \ref{fig:Affected-reurons.},
we denote the four vertices (faces) nearest to $\mathcal{V}$ as $\mathcal{V}_{1}$,
$\mathcal{V}_{2}$, $\mathcal{V}_{3}$, and $\mathcal{V}_{4}$ ($\mathcal{F}_{1}$,
$\mathcal{F}_{2}$, $\mathcal{F}_{3}$, and $\mathcal{F}_{4}$). Then
by using Eq. (\ref{eq:vetex0}), we have $\Gamma_{\mathcal{V}'}(\Xi)  =  \Gamma_{\mathcal{V}'}(\Xi;\sigma_{j}^{z}\rightarrow-\sigma_{j}^{z},\forall j\in\mathcal{V}),\;\text{for}\;\mathcal{V}'\neq\mathcal{V}_{1},\mathcal{V}_{2},\mathcal{V}_{3},\text{ or }\mathcal{V}_{4}$ and $\Gamma_{\mathcal{F}}(\Xi) =  \Gamma_{\mathcal{F}}(\Xi;\sigma_{j}^{z}\rightarrow-\sigma_{j}^{z},\forall j\in\mathcal{V}),\;\text{for}\;\mathcal{F}\neq\mathcal{F}_{1},\mathcal{F}_{2},\mathcal{F}_{3},\text{ or }\mathcal{F}_{4}.$
Canceling out these equal factors, Eq. (\ref{eq:Avphi}) reduces to
\begin{widetext}
\begin{eqnarray}
	\Gamma_{\mathcal{V}}(\Xi)\prod_{\mu=1}^{4}\Gamma_{\mathcal{V}_{\mu}}(\Xi)\Gamma_{\mathcal{F}_{\mu}}(\Xi) =  \Gamma_{\mathcal{V}}(\Xi;\sigma_{j}^{z}\rightarrow-\sigma_{j}^{z},\forall j\in\mathcal{V}) 
 \prod_{\mu=1}^{4}\Gamma_{\mathcal{V}_{\mu}}(\Xi;\sigma_{j}^{z}\rightarrow-\sigma_{j}^{z},\forall j\in\mathcal{V})\Gamma_{\mathcal{F}_{\mu}}(\Xi;\sigma_{j}^{z}\rightarrow-\sigma_{j}^{z},\forall j\in\mathcal{V}).\quad\quad\;\label{eq:VertEq}
\end{eqnarray}
\end{widetext}
Let $\varXi'_{\text{sub}}$ denote the spins of the corresponding
visible neurons belong to $\mathcal{V}_{1}$, $\mathcal{V}_{2}$,
$\mathcal{V}_{3}$, $\mathcal{V}_{4}$, $\mathcal{F}_{1}$, $\mathcal{F}_{2}$,
$\mathcal{F}_{3}$, or $\mathcal{F}_{4}$ (neurons in the shaded region
in Fig. \ref{fig:Affected-reurons.}). Eq. (\ref{eq:VertEq}) should
be satisfied for any configurations of $\varXi'_{\text{sub}}$, giving
a series of $2^{16}=65536$ equations. Directly solving these equations
is daunting. As discussed in the 1D case in the main text, we can
recast Eq. (\ref{eq:VertEq}) to an optimization problem and find
a solution numerically 
\begin{eqnarray*}
	b_{\mathcal{V}} & = & 0,\;W_{\mathcal{V};(\mathcal{V},\mu)}=\frac{i\pi}{2},\quad\forall\mu,\mathcal{V}.
\end{eqnarray*}
This gives the exact ANNQS representation of the 2D  toric
code state in the main text.

\renewcommand{\theequation}{B\arabic{equation}}
\setcounter{equation}{0}  
\renewcommand{\thefigure}{B\arabic{figure}}
\setcounter{figure}{0}  

\section*{APPENDIX B: EXCITED STATES WITH ABELIAN ANYONS}

\begin{figure}
	\includegraphics[width=0.486\textwidth]{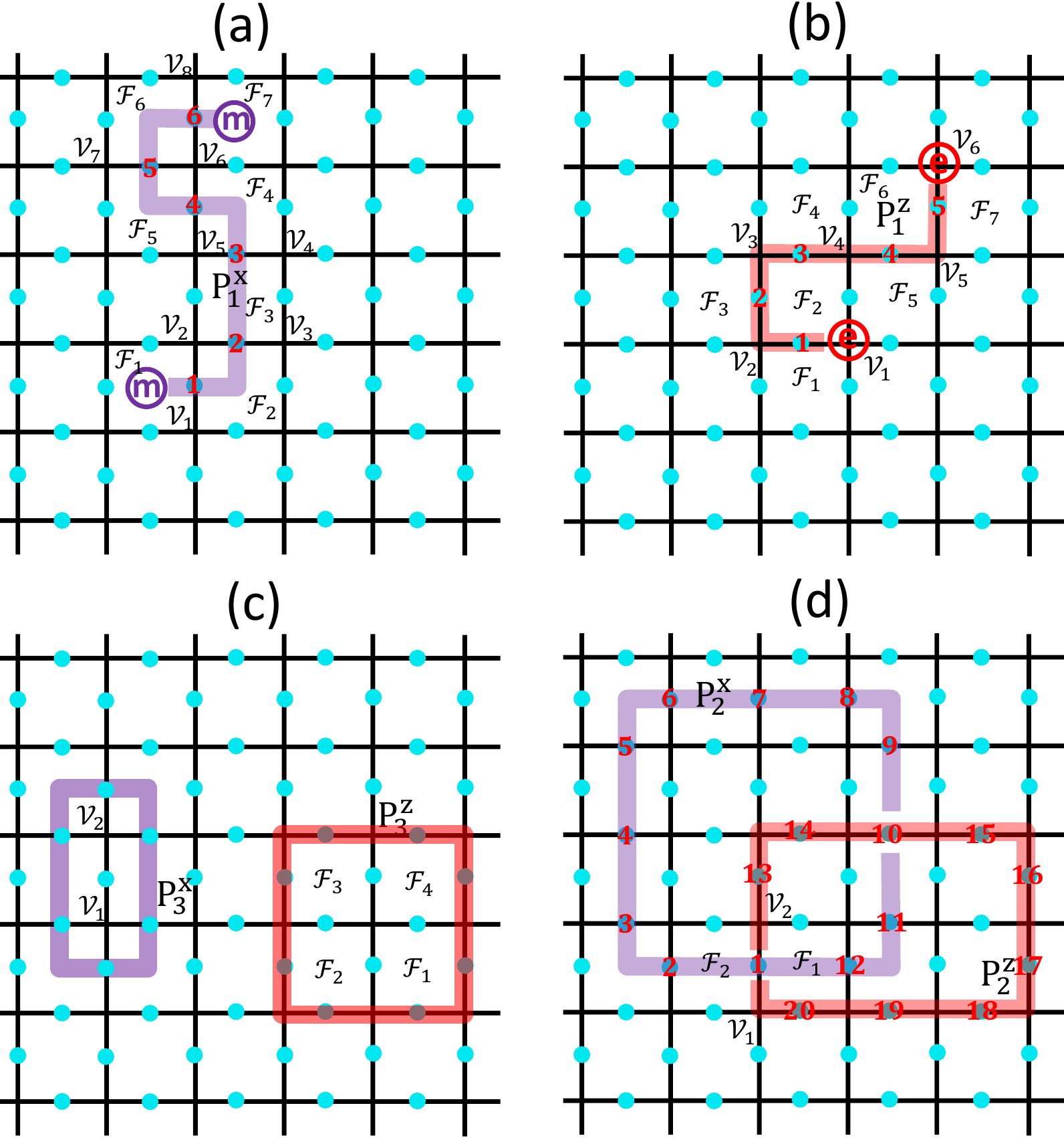}
	
	\caption{Anyons created by string operators and their mutual statistics.
	(a) Creating a pair of x-type quasiparticles living at faces $\mathcal{F}_1$ and $\mathcal{F}_7$ by apply string operator $S^x_{\text{P}^\text{x}_1}$ on the ground state $|G_{\text{toric}}^{\text{2D}}\rangle$. (b) Creating a pair of z-type quasiparticles by $S^z_{\text{P}^\text{z}_1}$. (c) A trivial contractible loop that leave the ground state unchanged.   (d) A Hopf link formed by two loops $\text{P}^\text{x}_2$ and $\text{P}^\text{z}_2$. Acting the corresponding string operators on the ground state will give rise to an overall phase $-1$, which manifests the nontrivial mutual statistics between the x- and z-type quasiparticles. 
	 \label{fig:anyons}}
\end{figure}

In this section, we provide more details on describing excited states with abelian anyons in the FRRBM framwork. We begin with an excited states with two x-type quasiparticles. As shown in Fig.\ref{fig:anyons}(a), we consider a string operator
\begin{eqnarray}
 S^x_{\text{P}^\text{x}_1}=\hat{\sigma}^x_6\hat{\sigma}^x_5\hat{\sigma}^x_4\hat{\sigma}^x_3\hat{\sigma}^x_2\hat{\sigma}^x_1.
  \end{eqnarray}  
When it acts on the ground state $|G_{\text{toric}}^{\text{2D}}\rangle$, anyons will be created and moved as follows \cite{kitaev2003fault,hu2009proposed}. First, acting $\hat{\sigma}^x_1$ will create a pair of x-type quasiparticles on faces $\mathcal{F}_1$ and $\mathcal{F}_2$. In the FRRBM language, it corresponds to flipping the signs of the weight parameters associated to the visible neuron at site $1$, i.e., 
$W_{\mathcal{V}_1;(\mathcal{V}_1,1)}\rightarrow -W_{\mathcal{V}_1;(\mathcal{V}_1,1)},\;W_{\mathcal{V}_2;(\mathcal{V}_2,1)}\rightarrow -W_{\mathcal{V}_2;(\mathcal{V}_2,1)},\;W_{\mathcal{F}_1;(\mathcal{F}_1,1)}\rightarrow -W_{\mathcal{F}_1;(\mathcal{F}_1,1)}$, and $W_{\mathcal{F}_2;(\mathcal{F}_2,1)}\rightarrow -W_{\mathcal{F}_2;(\mathcal{F}_2,1)}.$
To see that the x-type particles are located on faces $\mathcal{F}_1$ and $\mathcal{F}_2$, one can apply the face operators $B_{\mathcal{F}_1}$ and $B_{\mathcal{F}_2}$, to the new states $|\Psi^{(1)}\rangle=\hat{\sigma}^x_1|G_{\text{toric}}^{\text{2D}}\rangle$
\begin{eqnarray}
B_{\mathcal{F}_1}|\Psi^{(1)}\rangle&=&B_{\mathcal{F}_1}\hat{\sigma}^x_1|G_{\text{toric}}^{\text{2D}}\rangle \nonumber\\
&=&-\hat{\sigma}^x_1B_{\mathcal{F}_1}|G_{\text{toric}}^{\text{2D}}\rangle=-|\Psi^{(1)}\rangle,\label{eqBF1} \\ 
B_{\mathcal{F}_2}|\Psi^{(1)}\rangle&=&B_{\mathcal{F}_2}\hat{\sigma}^x_1|G_{\text{toric}}^{\text{2D}}\rangle \nonumber \\
&=&-\hat{\sigma}^x_1B_{\mathcal{F}_2}|G_{\text{toric}}^{\text{2D}}\rangle=-|\Psi^{(1)}\rangle. \label{eqBF2}
\end{eqnarray}
Thus, $|\Psi^{(1)}\rangle$ is an eigenstate of both $B_{\mathcal{F}_1}$ and $B_{\mathcal{F}_2}$ with eigenenergy $-1$, indicating a pair of local x-type quasiparticles at faces $\mathcal{F}_1$ and $\mathcal{F}_2$. For all other vertex or face operators, $|\Psi^{(1)}\rangle$ is an eigenstate with eigenenergy $1$ since they commute with $\hat{\sigma}^x_1$. This simply means no other excitation is created.  In the FRRBM context, Eqs.(\ref{eqBF1}-\ref{eqBF2}) can be verified by noting $\cos[\frac{\pi}{4}(\sum_{j\in \mathcal{F}_1,j\neq 1}\sigma^z_j-\sigma^z_1)]B_{\mathcal{F}_1}=-\cos[\frac{\pi}{4}(\sum_{j\in \mathcal{F}_1,j\neq 1}\sigma^z_j-\sigma^z_1)]$ and $\cos[\frac{\pi}{4}(\sum_{j\in \mathcal{F}_2,j\neq 1}\sigma^z_j-\sigma^z_1)]B_{\mathcal{F}_2}=-\cos[\frac{\pi}{4}(\sum_{j\in \mathcal{F}_2,j\neq 1}\sigma^z_j-\sigma^z_1)].$ 
Then we act $\sigma^x_2$ on $|\Psi^{(1)}\rangle$, and this create another pair of x-type particles on faces $\mathcal{F}_2$ and $\mathcal{F}_3$, but the x-type particle on face $\mathcal{F}_2$ will annihilate with the original one. Effectively, the original particle on face $\mathcal{F}_2$ is moved to face $\mathcal{F}_3$. In the FRRBM language, this process corresponds to flipping the signs of the weight parameters associated to the visible neuron at site 2, i.e.,
$W_{\mathcal{V}_2;(\mathcal{V}_2,2)}\rightarrow -W_{\mathcal{V}_2;(\mathcal{V}_2,2)},\;W_{\mathcal{V}_3;(\mathcal{V}_3,2)}\rightarrow -W_{\mathcal{V}_3;(\mathcal{V}_3,2)},\;W_{\mathcal{F}_2;(\mathcal{F}_2,2)}\rightarrow -W_{\mathcal{F}_2;(\mathcal{F}_2,2)}$, and$W_{\mathcal{F}_3;(\mathcal{F}_3,2)}\rightarrow -W_{\mathcal{F}_3;(\mathcal{F}_3,2)}$. 
This procedure continues until $\sigma^x_6$ has been applied and the x-type particle originally at face $\mathcal{F}_2$ will be moved to face $\mathcal{F}_7$. Correspondingly, the signs of all weight parameters associated to the visible neurons living on the path $\text{P}^x_1$ will be flipped. Thus, we obtained the exact RBM representation of $|\Psi_{\text{P}^\text{x}_1}\rangle$, as given in Eq. (11) in the main text.

In Fig.\ref{fig:anyons}(b), we consider a z-type string operator
\begin{eqnarray}
 S^z_{\text{P}^\text{z}_1}=\hat{\sigma}^z_5\hat{\sigma}^z_4\hat{\sigma}^z_3\hat{\sigma}^z_2\hat{\sigma}^z_1.
  \end{eqnarray}  
When we act it on the ground state $|G_{\text{toric}}^{\text{2D}}\rangle$, anyons will be created and moved as follows. First, acting $\hat{\sigma}^z_1$ will create a pair of z-type quasiparticles on vertexes $\mathcal{V}_1$ and $\mathcal{V}_2$. In the FRRBM language, it corresponds to adding a hidden neuron $h_1$ at site 1 with the additional nonzero parameter chosen to be 
\begin{eqnarray}
b_1=-\frac{i\pi}{2},\; W_{11}=\frac{i\pi}{2}. \label{parametervalue}
\end{eqnarray}
In other words, the hidden neuron  only connects to the visible neuron at site 1. To see that adding $h_1$ leads to the creation of two z-type quasiparticles on vertexes $\mathcal{V}_1$ and $\mathcal{V}_2$, one can apply the vertex operators $A_{\mathcal{V}_1}$ and $A_{\mathcal{V}_2}$, to the new states $|\varphi^{(1)}\rangle=\hat{\sigma}^z_1|G_{\text{toric}}^{\text{2D}}\rangle$. Due to the anticommutation relations, it is straightforward to show that $|\varphi^{(1)}\rangle$ is an eigenstate of both $A_{\mathcal{V}_1}$ and $A_{\mathcal{V}_2}$ with eigenenergy $-1$, giving a pair of  z-type quasiparticles at vertexes $\mathcal{V}_1$ and $\mathcal{V}_2$. In the FRRBM context, this can be obtained by noting $\cos[\frac{i\pi}{2}(-1+\sigma^z_1)]=-\cos[\frac{i\pi}{2}(-1-\sigma^z_1)]$ and $A_{\mathcal{V}_1}|G_{\text{toric}}^{\text{2D}}\rangle=A_{\mathcal{V}_2}|G_{\text{toric}}^{\text{2D}}\rangle=|G_{\text{toric}}^{\text{2D}}\rangle$.  We then act $\sigma^z_2$ on $|\varphi^{(1)}\rangle$ to effectively move the z-type particle originally on vertex $\mathcal{V}_2$ to vertex $\mathcal{V}_3$. This corresponds to adding another hidden neuron $h_2$ at site 2 with the same nonzero parameter as specified in Eq. (\ref{parametervalue}). We continue this procedure until $\sigma^z_5$ is applied and the z-type particle moved to vertex  $\mathcal{V}_6$. In the FRRBM context, five hidden neurons corresponding to the path $\text{P}^{\text{z}}_1$ will be added to represent the final state $|\varphi^{(5)}\rangle=S^z_{\text{P}^\text{z}_1}|G_{\text{toric}}^{\text{2D}}\rangle$. 

In Fig.\ref{fig:anyons}(c), we consider a closed x-type (z-type) string operator $S^x_{\text{P}^\text{x}_3}$ ($S^z_{\text{P}^\text{z}_3}$) defined on a contractible loop $\text{P}^{\text{x}}_3$ ($\text{P}^{\text{z}}_3$). It is obvious that  $S^x_{\text{P}^\text{x}_3}=A_{\mathcal{V}_1}A_{\mathcal{V}_2}$ and $S^z_{\text{P}^\text{z}_3}=B_{\mathcal{F}_1}B_{\mathcal{F}_2}B_{\mathcal{F}_3}B_{\mathcal{F}_4}$. Thus acting them on the ground state will leave it unaltered. In the FRRBM context, acting $S^x_{\text{P}^\text{x}_3}$ corresponds to flipping signs of all weight parameters associated to the visible neurons living on the path $\text{P}^x_3$. The new RBM state is the same as the original ground state, following from the facts that: (i) each face operator and $S^x_{\text{P}^\text{x}_3}$ correspond to either 0 or 2 common visible neurons; (ii) the sign flippings preserve the sign of the product of the two cosine factors related to $\mathcal{V}_1$ and $\mathcal{V}_2$. Similarly, acting $S^z_{\text{P}^\text{z}_3}$ corresponds to adding 8 hidden neurons living on $\text{P}^z_3$ and this will not alter the ground state, either.

In Fig.\ref{fig:anyons}(d), two string operators of different type form a Hopf link. in this case, anyons will be created, moved, and annihilated as follows. First, we act $\sigma^z_1$ on the ground state to create two z-type particles at $\mathcal{V}_1$ and $\mathcal{V}_2$, respective. We then act  $\sigma^x_1$ to create two x-type particles at $\mathcal{F}_1$ and $\mathcal{F}_2$, and $\sigma^x_{12}\sigma^x_{11}\sigma^x_{10}\sigma^x_9\sigma^x_8\sigma^x_7\sigma^x_6\sigma^x_5
\sigma^x_4\sigma^x_3\sigma^x_2$ to move one of the x-type particles along the path $\text{P}^x_2$ and then annihilate it with its partner. Finally, we act $\sigma^z_{20}\sigma^z_{19}\sigma^z_{18}\sigma^z_{17}\sigma^z_{16}
\sigma^z_{15}\sigma^z_{14}\sigma^z_{13}$ to move one of the z-type particles along the path $\text{P}^z_2$ and annihilate it with its partner. Effectively, a x-type particle is braided with a z-type particle and the ground state will gain a $-1$ phase factor due to the nontrivial mutual statistics \cite{kitaev2003fault}. In the FRRBM context, this process corresponds to the following steps: (i) adding a hidden neuron at site $1$ with parameters specified in Eq. (\ref{parametervalue}); (ii) flipping the signs of all weight parameters associated to path $\text{P}^x_2$; (iii) adding nine neurons along path $\text{P}^z_2$. Following similar reasoning as discussed above, it is straight forward to obtain that the final RBM state is the same as the ground state, but with an extra overall phase $-1$. This extra phase results from the beginning of the step (ii), noting that flip the sign of $W_{11}$ leads to an extra $-1$ factor.

\bibliographystyle{apsrev4-1-title}

\bibliography{Dengbib}

\end{document}